\documentclass{emulateapj}
\usepackage{amsmath}
\usepackage{color}
\usepackage{ulem}

\shorttitle{The VMC Survey XXIV. Hierarchical Star Formation}
\shortauthors{N.-C. Sun et al.}
\slugcomment{Accepted for publication in The Astrophysical Journal}

\begin{document}

\title{The VMC Survey XXII. Hierarchical Star Formation in the 30~Doradus-N158-N159-N160 Star-Forming Complex}

\author{Ning-Chen~Sun$^{1,2}$,
Richard~de~Grijs$^{1,2,3}$,
Smitha~Subramanian$^{1}$,
Maria-Rosa~L.~Cioni$^{4, 5, 6}$,
Stefano~Rubele$^{7, 8}$,
Kenji~Bekki$^{9}$,
Valentin~D.~Ivanov$^{10, 11}$,
Andr\'es~E.~Piatti$^{12, 13}$,
Vincenzo~Ripepi$^{14}$}
\affil{$^1$Kavli Institute for Astronomy and Astrophysics, Peking University, Yi He Yuan Lu 5, Hai Dian District, Beijing 100871, China; sunnc@foxmail.com, grijs@pku.edu.cn \\
$^2$Department of Astronomy, Peking University, Yi He Yuan Lu 5, Hai Dian District, Beijing 100871, China \\
$^3$International Space Science Institute -- Beijing, 1 Nanertiao, Hai Dian District, Beijing 100190, China \\
$^4$Universit\"at Potsdam, Institut f\"ur Physik und Astronomie, Karl-Liebknecht-Str. 24/25, Potsdam 14476, Germany \\
$^5$Leibniz-Institut f\"ur Astrophysik Potsdam, An der Sternwarte 16, Potsdam 14482, Germany \\
$^6$University of Hertfordshire, Physics, Astronomy and Mathematics, College Lane, Hatfield AL10 9AB, UK \\
$^7$Osservatorio Astronomico di Padova -- INAF, vicolo dell'Osservatorio 5, Padova I-35122, Italy \\
$^8$Dipartimento di Fisica e Astronomia, Universit\`a di Padova, vicolo dell'Osservatorio 2, Padova I-35122, Italy \\
$^9$ICRAR, M468, The University of Western Astralia 35 Stirling Highway, Crawley Western Astralia, 6009, Australia \\
$^{10}$ESO European Southern Observatory, Ave. Alonso de Cordova 3107, Casilla 19, Chile \\
$^{11}$ESO Garching: ESO,  Karl-Schwarzschild-Str. 2, 85748 Garching bei M\"unchen, Germany  \\ 
$^{12}$Observatorio Astron\'omico, Universidad Nacional de C\'ordoba, Laprida 854, 5000, C\'ordoba, Argentina \\
$^{13}$Consejo Nacional de Investigaciones Cient\'ificas y T\'ecnicas, Av. Rivadavia 1917, C1033AAJ Buenos Aires, Argentina \\
$^{14}$INAF -- Osservatorio Astronomico di Capodimonte, Via Moiariello 16, I-80131 Naples, Italy}

\begin{abstract}
We study the hierarchical stellar structures in a $\sim$1.5 deg$^2$ area covering the 30~Doradus-N158-N159-N160 star-forming complex with the VISTA Survey of the Magellanic Clouds. Based on the young upper main-sequence stars, we find that the surface densities cover a wide range of values, from log($\Sigma\cdot$pc$^2$)~$\lesssim$~$-$2.0 to log($\Sigma\cdot$pc$^2$)~$\gtrsim$~0.0. Their distributions are highly non-uniform, showing groups that frequently have sub-groups inside. The sizes of the stellar groups do not exhibit characteristic values, and range continuously from several parsecs to more than 100~pc; the cumulative size distribution can be well described by a single power law, with the power-law index indicating a projected fractal dimension $D_2$~=~1.6~$\pm$~0.3. We suggest that the phenomena revealed here support a scenario of hierarchical star formation. Comparisons with other star-forming regions and galaxies are also discussed.
\end{abstract}


\section{Introduction}

It has been suggested that star formation is hierarchical as revealed by stellar structures spanning a wide range of scales, from large stellar complexes and aggregates to small associations and clusters \citep{Efremov1995, Elmegreen2000}. Hierarchical stellar structures have been investigated for a few famous star-forming regions, such as Taurus, Orion, Ophiuchus, etc. in the Milky Way \citep{Gomez1993, Larson1995, Simon1997} and NGC~345 in the Small Magellanic Cloud \citep[SMC;][]{Schmeja2009, Gouliermis2014}. On larger scales, a number of galaxies display galaxy-wide hierarchies formed by young stars \citep[e.g.][]{Bastian2007, Bastian2009, Gieles2008, Elmegreen2001, Elmegreen2006, Elmegreen2014, Gouliermis2010, Gouliermis2015}. These studies show that hierarchical stellar structures, significant for a high degree of substructure and an absence of characteristic scales, exhibit self-similar and fractal properties. These features are also found for the structures formed by the interstellar medium (ISM), such as clumps and filaments on various scales \citep[e.g.][]{Elmegreen1996, RomanDuval2010}. This similarity suggests that the hierarchical structures of the young stars may originate from those of their parental ISM, which are in turn related to turbulence \citep{Larson1981, Elmegreen2008, Federrath2009, Girichidis2012}, agglomeration with fragmentation \citep{Carlberg1990, McLaughlin1996}, and self-gravity (\citealt{deVega1996}; see also the review of \citealt{Elmegreen2000}). After birth, the hierarchical stellar structures evolve toward uniform distributions, with the substructures becoming eliminated in roughly one crossing time \citep{Gieles2008, Bastian2009, Gouliermis2015}.

30~Doradus (30~Dor hereafter, also known as the Tarantula Nebula) in the Large Magellanic Cloud (LMC), with R.A.~(J2000)~=~05$^{\rm h}$38$^{\rm m}$38$^{\rm s}$, Dec.~(J2000)~=~$-$69$^{\rm o}$05$\arcmin$42$\arcsec$, is among the most famous astronomical objects. It is by far the most luminous and massive star-forming region in the Local Group \citep{Kennicutt1986} and hosts at its center the star cluster NGC~2070, which in turn is a collective of dense subclusters \citep{Walborn1997, Sabbi2012}. On the other hand, three active star-forming regions, N158, N159, and N160, are found to the south of 30~Dor \citep{Nakajima2005, Galametz2013}. They are located in the northern part of the LMC's remarkable ``molecular ridge", which stretches $\sim$2~kpc in a nearly north-south straight line and contains almost one third of the LMC's molecular gas traced by CO \citep{Cohen1988, Kutner1997, Johansson1998, Fukui2008, Ott2008}. Detailed studies of the hierarchical stellar structures in these star-forming regions have not been reported. Although \citet{Bastian2009} have investigated this issue for the entire LMC, their stellar sample did not reach the necessary resolution to reveal the substructures of these regions. This paper thus aims to carry out such a study for the 30~Dor-N158-N159-N160 star-forming complex based on young upper main-sequence (UMS) stars, which have been carefully selected from the VISTA Survey of the Magellanic Clouds \citep[VMC;][]{Cioni2011}. We use dendrograms to illustrate the ``parent--child" relations of star groups and subgroups identified on various density levels; and additionally the group size distribution is explored. This will demonstrate the hierarchical structures formed by the young UMS stars, which may indicate a scenario of hierarchical star formation in this complex.

This paper is organized as follows. Section~\ref{data.sec} describes the VMC survey and the data used in this work. Selection of UMS stars is outlined in Section~\ref{sample.sec}, while in Section~\ref{distribution.sec} we show their spatial distributions. We identify groups of UMS stars and reveal their ``parent--child" relations using dendrograms in Section~\ref{dendro.sec}, and the group sizes are investigated in Section~\ref{size.sec}. We finally close this paper with a summary and conclusions.

\section{The VMC Survey and Data}
\label{data.sec}
Data used in this work come from the VMC survey \citep{Cioni2011}, which is carried out with the 4.1~m Visible and Infrared Survey Telescope for Astronomy \citep[VISTA;][]{Sutherland2015}. The VMC survey is a multi-epoch, uniform, and homogeneous photometric survey of the Magellanic system performed in the near-infrared $Y$, $J$, and $K_{\rm s}$ filters (centered at 1.02 $\mu$m, 1.25 $\mu$m, and 2.15 $\mu$m, respectively). On completion, the survey is expected to cover $\sim$170~deg$^2$ of the Magellanic system. A sequence of six offset positions, each one called a $\textit{paw-print}$, is used to fill in the gaps between the 16 detectors of the VISTA infrared camera \citep[VIRCAM;][]{Dalton2006}; the combined image, or a VISTA $tile$, covers an area of $\sim$1.5 deg$^2$. Each tile in the VMC survey is designed to be observed at 3 epochs in the $Y$ and $J$ bands, and 12 epochs in the $K_{\rm s}$ band. The total exposure times are 2400 s in the $Y$ and $J$ bands, and 9000 s in the $K_{\rm s}$ band for most regions of each tile, except for the tile edges and some areas of extra overlap among the detectors. The detectors usually have saturation limits of $Y$ = 12.9 mag, $J$ = 12.7 mag, and $K_{\rm s}$ = 11.4 mag, and the stacked images from all epochs can provide sources with typical 5$\sigma$ limiting magnitudes of $Y$ = 21.9 mag, $J$ = 22.0 mag, and $K_{\rm s}$ = 21.5 mag; but the saturation limits and photometric depths also depend on sky conditions and crowding \citep{Cioni2011, Tatton2013}. This depth reaches the oldest main-sequence (MS) turn-offs in both the LMC and SMC \citep{Kerber2009}.

The 30~Dor-N158-N159-N160 star-forming complex is analyzed based on tile LMC~6\_6 (see \citealt{Cioni2011} for tile definitions). We retrieved the data of tile LMC~6\_6 as part of VMC Data Release 3 from the VISTA Science Archive (VSA). The details of the VSA and the VISTA data flow pipeline can be found in \citet{Cross2012} and \citet{Irwin2004}, respectively. We have used point-spread-function (PSF) photometry from \citet{Rubele2012} instead of aperture photometry to reduce the influence of source crowding \citep{Tatton2013}. \citet{Rubele2012} generated deep, PSF-homogenized tiles by stacking images from different epochs, each one formed by the combination of six pawprints. They performed PSF photometry on the deep tiles and estimated the photometric errors and local completeness using artificial star tests. Specifically, the local completeness is estimated in a ring of radius 0.025~deg around each source and in bins of $\pm$0.05~mag separately in the $Y$, $J$, and $K_{\rm s}$ bands. We use their local completeness estimates to assign weights to stars in our sample (see Sections~\ref{sample.sec} and \ref{distribution.sec}). A detailed description of the PSF homogeneity and photometry can also be found in \citet{Rubele2015}.

The top half of VISTA's detector 16 has varying quantum efficiency on short timescales, making accurate flat fielding impossible. This leads to worse signal-to-noise ratios and unreliable data in regions covered by that detector \citep{Rubele2012, Tatton2013}. For tile LMC~6\_6, the region affected is the southwest corner with R.A.~(J2000)~$<$~05$^{\rm h}$33$^{\rm m}$55$^{\rm s}$ and Dec.~(J2000)~$<$~$-$69$^{\rm o}$43$\arcmin$48$\arcsec$. This region is excluded from our analysis.

\section{The Upper Main Sequence Sample}
\label{sample.sec}

\begin{figure*}
\centering
\includegraphics[scale=0.80,angle=0]{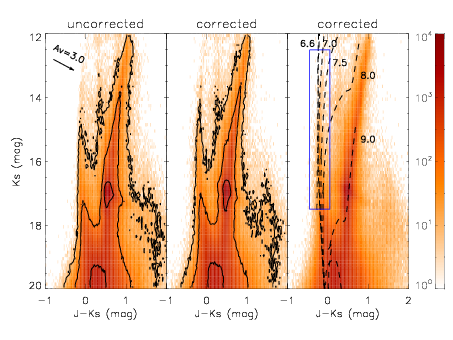}
\caption{($J-K_{\rm s}$,~$K_{\rm s}$) color--magnitude Hess diagrams of stars in tile LMC~6\_6. In the left-hand panel, interstellar extinction has not been corrected for, while in the middle and right-hand panels, interstellar extinction has been corrected for according to \citet{Tatton2013}. The color scales in all three panels show the number of stars in each color--magnitude bin; the bin size is 0.02~mag in color and 0.10~mag in magnitude. The contours in the left-hand and middle panels correspond to levels of 10$^1$, 10$^2$, and 10$^3$ stars per bin from the outside to the inside. The arrow in the left-hand panel shows the reddening vector corresponding to $A_V$~=~3.0~mag. In the right-hand panel, the blue solid box shows the selection criterion of UMS stars and dashed lines are the MS to RGB sections of PARSEC (version 1.2S) isochrones of metallicity $Z$~=~0.008 and ages log($\tau$/yr)~=~6.6, 7.0, 7.5, 8.0, and 9.0, shifted by a distance modulus of ($m-M$)$_0$~=~18.49~mag. Offsets of 0.026~mag in $J$ and 0.003~mag in $K_{\rm s}$,  given by \citet{Rubele2015}, have been subtracted from the isochrone magnitudes to correct for the small differences between the VISTA system and the model Vega system. The color scale in the right-hand panel is the same as that in the middle panel; we overplot the contours, sample selection criterion, and isochrones in the different panels simply for clarity.}
\label{cmd.fig}
\end{figure*}

The left-hand panel of Fig.~\ref{cmd.fig} shows the ($J-K_{\rm s}$,~$K_{\rm s}$) CMD of sources in tile LMC~6\_6 without correcting for extinction. The MS, red-giant branch (RGB), and red clump (RC) populations are all resolved. We use the ($J-K_{\rm s}$) color instead of ($Y-K_{\rm s}$) because the former suffers less from interstellar extinction. Still, effects of interstellar extinction varying across the tile are present, evidenced by the width of the MS and the RGB, and the elongated tail associated with the RC toward fainter magnitudes and redder colors.

We estimate the extinction by taking advantage of the extinction values of $\sim$1.5 $\times$ 10$^5$ RC stars in this tile provided by \citet{Tatton2013}. We bin both all stars and their RC stars into spatial grids with a grid size of 1\arcmin~$\times$~1\arcmin; the median extinction value of RC stars is taken as the extinction estimate for all stars in each bin. We use the extinction coefficients,  $A_J$~=~0.283~$\times A_V$ and $A_{K_{\rm s}}$~=~0.114~$\times A_V$, computed from the \citet{Cardelli1989} extinction curve with $R_{\rm V}$~=~3.1 \citep{Girardi2008}. \citet{DeMarchi2016} reported a new extinction law for 30~Dor, characterized by $R_{V}$~$\sim$~4.5. However, using this law would not lead to significant changes in the corrections for the near-infrared wavelengths considered in this paper \citep{Gordon2003}. Contamination by RGB stars is not important, since their intrinsic color exhibits only a small difference of $\sim$0.05$\pm$0.02~mag from that of the RC stars \citep{Tatton2013}, thus introducing very minor, if any, errors in correcting the CMD. The middle panel of Fig.~\ref{cmd.fig} shows the CMD after correcting for extinction. The MS and RGB become tighter and the elongated tail associated with the RC appears less prominent.

The extinction in the LMC is subject to a population dependence, which is, however, difficult to model \citep{Zaritsky2004, Cignoni2015}. Compared with intermediate-age RC stars, the young stars form in gas-rich regions with high extinction, but their ionizing fluxes and stellar winds tend to reduce extinction by evacuating the surrounding material. The problem is further compounded by their different spatial distributions with respect to the dust. Along the line of sight, dust may lie in the foreground and/or background of a star, and only the foreground dust contributes to its extinction. Perpendicular to the line of sight, dust in star-forming regions also exhibits a high degree of spatial variability \citep[e.g.][]{Lombardi2010, Lombardi2011, Cambresy2013}. This is also demonstrated by the \textit{Herschel} dust emission map across the SMC star-forming region NGC~345 (also known as N66), which is highly variable on very small scales and rich in structures of knots, arcs and filaments, etc. \citep{Hony2015}. It is very challenging and beyond the scope of this work to take into account all these considerations by obtaining a population-dependent, three-dimensional extinction map with high angular resolution. Still, our treatment of extinction is reasonable as the features become more well-defined in the extinction-corrected CMD.

In the right-hand panel of Fig.~\ref{cmd.fig}, we overplot PARSEC isochrones \citep[version 1.2S;][]{Bressan2012} of metallicity $Z$~=~0.008 and ages log($\tau$/yr)~=~6.6, 7.0, 7.5, 8.0, and 9.0, shifted by an LMC distance modulus of ($m-M$)$_0$~=~18.49~mag \citep{Pietrzynski2013, deGrijs2014} and assuming zero extinction. We have subtracted offsets of 0.026~mag in $J$ and 0.003~mag in $K_{\rm s}$, given by \citet{Rubele2015}, from the isochrones to correct for the small differences between the VISTA system and the model Vega system. The differences are caused by the unaccounted for non-linearities in the color--color relation used to calibrate VISTA's photometric zeropoint with 2MASS stars \citep[for details, see][]{Rubele2012, Rubele2015}. It is clear that the upper part of the MS corresponds to a very young population, since it coincides well with isochrones of log($\tau$/yr)~=~6.6 and 7.0. We thus select a sample of UMS stars using color and magnitude cuts, $-$0.45~$\le$~$J-K_{\rm s}$~$\le$~0.05~mag and 12.5~$\le$~$K_{\rm s}$~$\le$~17.5~mag, which are shown as the blue solid-line box in the right-hand panel of Fig.~\ref{cmd.fig}. This sample contains $\sim$1.5~$\times$~10$^4$ stars.

\begin{figure*}
\centering
\includegraphics[scale=0.65,angle=0]{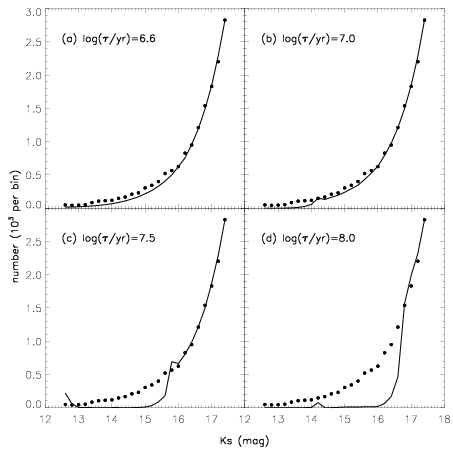}
\caption{$K_{\rm s}$-band luminosity functions of the observed UMS sample (dots) and synthetic ones (lines) with ages log($\tau$/yr)~=~6.6, 7.0, 7.5, and 8.0 (see the text for details). The observed UMS sample has been corrected for interstellar extinction and local completeness. The luminosity functions of the synthetic samples are normalized so that they have equal numbers of stars in the faintest magnitude bin as the observed sample. The two younger synthetic UMS samples have luminosity functions quite similar to the observed one, while the two older synthetic UMS samples have significantly fewer bright stars because they have evolved away from the MS. The Poissonian error bars of the observed luminosity function are smaller than or comparable to the symbol size.}
\label{Ks.fig}
\end{figure*}

We confirm that the UMS stars in our sample comprise indeed a young population by comparing their $K_{\rm s}$-band luminosity function with synthetic ones (Fig.~\ref{Ks.fig}). For the observed UMS sample, we take into account the photometric completeness by assigning to each star a weight $w$~=~1/$f$, where $f$ is the local completeness; because we are using the $J$ and $K_{\rm s}$ bands, $f$ is estimated from the lowest value of the $J$- and $K_{\rm s}$-band local completenesses, $f_J$ and $f_{K_{\rm s}}$, as estimated by \citeauthor{Rubele2012} (\citeyear{Rubele2012}, \citeyear{Rubele2015}; Section~\ref{data.sec}). On the other hand, we create four synthetic populations of metallicity $Z$~=~0.008 and ages log($\tau$/yr)~=~6.6, 7.0, 7.5, and 8.0. We assume a \citet{Chabrier2001} stellar initial mass function (IMF), a 30\% binary fraction, and that the binaries are non-interacting systems with primary/secondary mass ratios evenly distributed from 0.7 to 1.0 \citep{Kerber2009, Rubele2012, Rubele2015}. In each synthetic population, sufficiently large numbers of stars are created and placed at the LMC's distance modulus of 18.49~mag \citep{Pietrzynski2013, deGrijs2014}. Synthetic UMS samples are obtained by applying the same selection criterion to the synthetic populations as to the real data; their luminosity functions are then calculated.

Figure~\ref{Ks.fig} shows that the luminosity functions of the log($\tau$/yr)~=~6.6 synthetic UMS sample and the observed one are consistent, although the latter seems to contain slightly more stars with $K_{\rm s}$~$<$~16.0~mag (Fig.~\ref{Ks.fig}a); the log($\tau$/yr)~=~7.0 synthetic sample also has a very similar luminosity function, except for a slight lack of stars brighter than $K_{\rm s}$~=~14.0~mag (Fig.~\ref{Ks.fig}b). Synthetic samples with log($\tau$/yr)~=~7.5 and 8.0 deviate from the observed sample even more significantly, in that they have many fewer stars brighter than $K_{\rm s}$~=~15.7~mag or 16.8~mag, respectively (Fig.~\ref{Ks.fig}c and \ref{Ks.fig}d). The comparison leads us to suggest that the observed UMS sample is indeed composed of very young stars with ages of a few million years.

The VLT-FLAMES Tarantula Survey \citep[VFTS;][]{Evans2011} has collected fiber spectroscopy of about 800 early-type stars in a 25$\arcmin$ diameter field of the 30~Dor region, and spectroscopic classification of 352 O--B0 and 438 B-type stars has been done by \citet{Walborn2014} and \citet{Evans2015}, respectively. We compare our UMS sample with their OB-type stars; 279 O-type and 313 B-type stars observed by VFTS are covered by our UMS sample, and their spectral types range from O2 (earliest) to B5 (latest). This comparison again indicates the youth of the UMS sample.

Contamination by foreground Galactic stars and background galaxies is minimal. The former reside at typically ($J - K_{\rm s}$)~=~0.7~mag, and background galaxies have colors even redder than this value. Thus, they do not overlap with the UMS stars in the CMD and are not included by the color--magnitude selection criterion.

\section{Spatial Distributions}
\label{distribution.sec}

\begin{figure*}
\center
\begin{tabular}{cc}
\includegraphics[scale=0.75,angle=0]{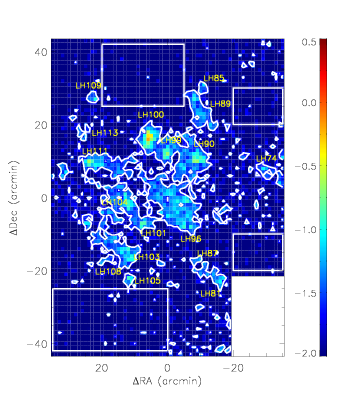}
\includegraphics[scale=0.75,angle=0]{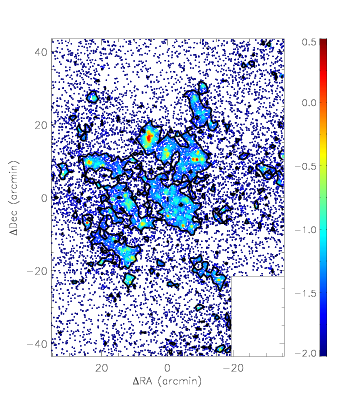}
\end{tabular}
\caption{Surface-density map of the UMS sample. In the left-hand panel, surface densities are derived from simple star counts in grids of 1$\arcmin \times$~1$\arcmin$, while in the right-hand panel, surface densities are calculated locally around each UMS star with the $n^{\rm th}$ nearest-neighbor method ($n$~=~10; see the text for details). Surface-density values are shown as a color scale, according to the color bar in units of logarithmic number of stars pc$^{-2}$. The four white boxes in the left-hand panel are the regions where we estimate the LMC background, and the level corresponding to 3$\sigma$ above the average of the background is overlaid as white and black contours in the left- and right-hand panels, respectively. The contours are computed only from the surface densities derived with simple star counts (i.e., the left-hand panel), but the same contours are also overlaid on the right-hand panel to indicate the large-scale stellar distributions. The yellow annotations in the left-hand panel label the associations catalogued by \citet{Lucke1970}. The (0, 0) position corresponds to R.A.~(J2000)~=~05$^{\rm h}$37$^{\rm m}$40$^{\rm s}$, Dec.~(J2000)~=~$-$69$^{\rm o}$22$\arcmin$18$\arcsec$. The southwest blank corner corresponds to the region affected by worse signal-to-noise ratios associated with detector 16.}
\label{density.fig}
\end{figure*}

\citet{Schmeja2011} compared different algorithms in constructing stellar density maps and identifying star clusters as overdensities; he showed that clusters with low overdensities or highly hierarchical structure are only reliably identified by methods with inherent smoothing, such as simple star counts and the $n^{\rm th}$ nearest-neighbor method \citep{Casertano1985, Megeath2016}. Thus, in this section we apply these two methods to the UMS sample to estimate their surface densities; the surface-density map derived with the $n^{\rm th}$ nearest-neighbor method is then used to identify stellar groups in Section~\ref{dendro.sec}. In the first method, we use a spatial grid of 1$\arcmin \times $1$\arcmin$ cells and determine the total weight of stars in each bin, with each star assigned a weight in the same way as described in Section~\ref{sample.sec}. In the second method, surface densities are calculated around each star with
\begin{equation}
\Sigma_n = \dfrac{w_{\rm tot}}{\pi r_n^2}\ ,
\end{equation}
where $r_n$ is the distance to the $n^{\rm th}$ nearest neighbor, and $w_{\rm tot}$ is the sum of the weights of all $n$+1 UMS stars at a radius of $r \le r_n$:
\begin{equation}
w_{\rm tot} = (w_0 - 1) + \sum^{n-1}_{i=1} w_i + 0.5 \times (w_n - 1)\ .
\end{equation}
\citet{Casertano1985} argued that the $n^{\rm th}$ nearest-neighbor method provides an unbiased estimate of density if neither the central star nor its $n^{\rm th}$ nearest neighbor are counted. In other words, we take into account the total weight inside the annulus between the central star and its $n^{\rm th}$ nearest neighbor. Thus, unity is subtracted from the central star, and its excess weight above unity, i.e. $w_0 - 1$, represents the over-density inside the annulus; similarly, 1 is subtracted from its $n^{\rm th}$ nearest neighbor, and half of its excess weight above 1, i.e. 0.5~$\times$~($w_n - 1$), is assumed to be inside the annulus. We use $n$~=~10, which has been shown to be a reasonable choice to balance statistical noise and locality \citep{Gouliermis2010, Megeath2016}. The separations between stars and their 10$^{\rm th}$ nearest neighbors range from 0.1$\arcmin$ in the densest area, to 3.6$\arcmin$ in low-surface-density regions, with a median value of 0.9$\arcmin$. Surface-density maps derived with the two methods are shown in Fig.~\ref{density.fig}.

Figure~\ref{density.fig} shows that the surface densities are generally low in the outer part of the tile. The UMS stars in the low-surface-density areas could come from the LMC's general background field, which are not associated with the star-forming complex. As we will show later in this paper, the young stars are hierarchically grouped in a way that small and compact stellar structures reside in larger and looser ones. Thus, there is also a possibility that the UMS stars in the low-surface-density areas belong to a larger stellar structure created in the star-forming process, but our sky coverage is not large enough to reveal it. In this work we do not try to reach scales larger than a single tile, thus we characterize this low surface density only from the LMC's background contribution. This should not make significant differences to our results, considering its very small values compared with the higher surface densities of the star-forming complex. We estimate the background level using four subregions, which are devoid of surface-density enhancements and outlined with white boxes in the left-hand panel of Fig.~\ref{density.fig}. We calculate the average and standard deviation of the surface densities derived with simple star counts of the bins in the subregions, iteratively rejecting 3$\sigma$ outliers. The average value is log($\overline{\Sigma}\cdot$pc$^2$)~=~$-$2.30 and the standard deviation is log($\sigma\Sigma\cdot$pc$^2$)~=~$-$2.28. The former is used as a measure of the background level, and the latter is an estimate of its fluctuations.

High surface densities are found in the central part of the tile, with roughly $-$20$\arcmin \le \Delta$R.A.~$\le$~25$\arcmin$,  $-$20$\arcmin \le \Delta$Dec.~$\le$~30$\arcmin$. The peak surface density, which occurs in the center of 30 Dor with $\Delta$R.A.~=~6$\arcmin$ and $\Delta$Dec.~=~17$\arcmin$, reaches log($\Sigma\cdot$pc$^2$)~$\sim$~$-$0.20, derived from simple star counts, or log($\Sigma\cdot$pc$^2$)~$\sim$~0.25, obtained with the $n^{\rm th}$ nearest-neighbor method, where $n$~=~10. The discrepancies between results obtained with the two methods arise from their different smoothing scales. The star count method has a fixed smoothing scale of 1$\arcmin$; the $n^{\rm th}$ nearest-neighbor method, in contrast, has variable smoothing scales, which are the separations between stars and their $n^{\rm th}$ nearest neighbors. As previously mentioned, the separations range from 0.1$\arcmin$ to 3.6$\arcmin$ for $n$~=~10, depending on the surface densities.

Also note that the majority of UMS stars are distributed in a number of highly fractured groups. The largest ones, with scales of several to $\sim$20$\arcmin$, are consistent with the OB associations catalogued by \citet{Lucke1970}, whose designations have been labeled in the figure. Here we recall that it can be difficult to define clusters, associations, or larger structures, and equally difficult to make clear distinctions between them \citep{Bressert2010, Krumholz2014}. \citet{Gieles2011} suggested that bound and unbound stellar systems can be distinguished by comparing the ages and crossing times; yet observationally it is not trivial to obtain these parameters. In order to avoid ambiguity, we will refer to any stellar structure with a surface density enhancement as a group, regardless of its size or dynamical state; the critical surface density to define an ``enhancement" is not fixed but can be varied arbitrarily, and by doing this we will reveal the hierarchy of stellar groups on a range of scales in Sections~\ref{dendro.sec} and \ref{size.sec}. But here we temporarily set the critical surface density at 3$\sigma$ above the average of the background level, so that we can investigate the stellar distributions on large scales. This corresponds to log($\Sigma_{\rm cr}\cdot$pc$^2$)~$\sim$~$-$1.7, which is shown as white or black contours in both panels of Fig.~\ref{density.fig}.

Most prominently in Fig.~\ref{density.fig}, the 30~Dor Nebula contains three populous groups, LH~100, LH~99, and LH~90 from its center to the west. In the vicinity of 30~Dor, there are three large, diffuse stellar groups, LH~89, LH~111, and LH~96, to its northwest, east, and south, respectively. In the southeastern part of the tile, there are another few groups with significant stellar overdensities, i.e., LH~101, LH~104, LH~103, and LH~105, which correspond to the N158, N160, and N159 star-forming regions \citep{Bolatto2000, Nakajima2005, Galametz2013}. A number of smaller groups can also be seen across the tile, e.g. LH~74 in the west, and even more that have not been catalogued by \citet{Lucke1970}.

Many of the groups described above are, in turn, highly sub-grouped. For example, LH~96 is far from centrally concentrated and contains multiple surface density peaks. This has been noted by \citet{Lucke1970}, who identify four smaller associations inside, i.e. LH~93, LH~94, LH~97, and LH~98. Similarly, the star-forming region, N158, contains two surface density peaks, LH~101 and LH~104, as catalogued by \citet{Lucke1970} and also revealed here. Although not obvious in Fig.~\ref{density.fig}, LH~100 contains two star clusters, NGC~2070 at the center of 30~Dor and Hodge~301 slightly to the northwest \citep{Cignoni2015}; the former is, in turn, a collective of smaller star clusters \citep{Walborn1997, Sabbi2012}. A statistical analysis of the hierarchical structures is given below.

\section{Group Identification and Dendrograms}
\label{dendro.sec}

In this section, we investigate the hierarchy of UMS star groups through the so-called $dendrograms$ \citep{Houlahan1992, Rosolowsky2008, Gouliermis2010}. Dendrograms are structure trees showing the ``parent--child" relations of groups with surface density enhancements over different critical values. We vary the critical surface density from log($\Sigma_{\rm cr}\cdot$pc$^2$)~=~$-$1.7 to $-$0.1; the lower limit corresponds to 3$\sigma$ above average of the background level, and the upper limit is chosen close to the peak surface density (see Section~\ref{distribution.sec}). Between them we define levels of critical surface densities with equal logarithmic steps of 0.2~dex. This step is arbitrarily chosen in order to avoid too many levels in displaying the dendrograms, considering that the surface densities span two orders of magnitude; reducing this step and increasing the number of levels would not change the conclusion reached in this section. On each level we use a friends-of-friends group identification algorithm \citep{Megeath2016}. To identify stellar groups, we use the surface densities estimated with the $n^{\rm th}$ nearest-neighbor method ($n$~=~10; Section~\ref{distribution.sec}). For any star with $\Sigma_n$~$\ge$~$\Sigma_{\rm cr}$, its 10 nearest neighbors which also show $\Sigma_n$~$\ge$~$\Sigma_{\rm cr}$ are its friends; friends of its friends are also friends, and so on. To avoid spurious detections, we require additionally that each group should contain at least $N_{\rm min}^\star$ UMS stars. Smaller values of $N_{\rm min}^\star$ would lead to more spurious detections, but larger values of $N_{\rm min}^\star$ would miss small groups which are not massive enough to contain sufficient number of UMS stars at the high-mass end of their IMF. As we shall see in Section~\ref{size.sec}, this effect is significant for groups of sizes smaller than 8~pc. In this section, however, this effect does not affect the conclusions  reached; thus we arbitrarily set $N_{\rm min}$~=~10 in the following analysis. The group identification process is repeated on each level of critical surface density, and we keep track of the connections between ``parent" groups on lower levels and ``child" groups on higher levels.

\begin{figure*}
\center
\begin{tabular}{cc}
\includegraphics[scale=0.75,angle=0]{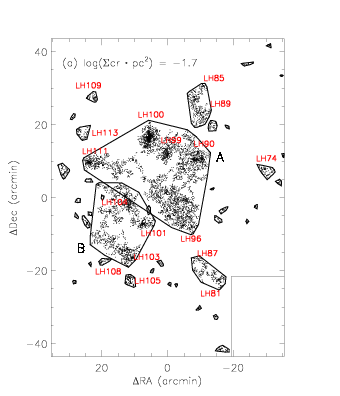}
\includegraphics[scale=0.75,angle=0]{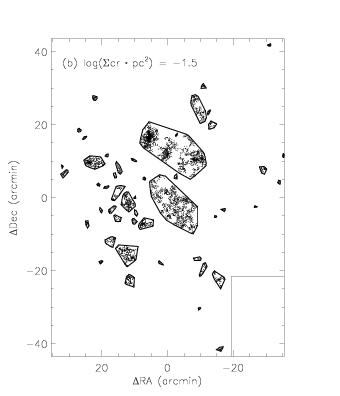} \\
\includegraphics[scale=0.75,angle=0]{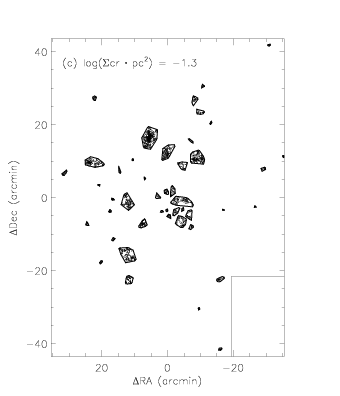}
\includegraphics[scale=0.75,angle=0]{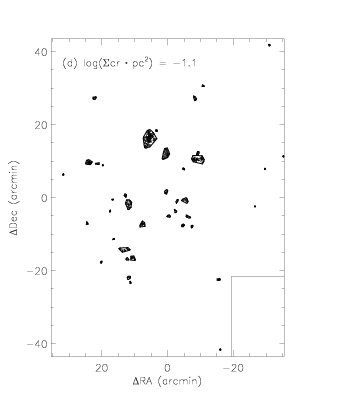}\\
\end{tabular}
\caption{Groups of UMS stars identified with $N_{\rm min}^\star$~=~10 and different critical surface densities labeled in each panel (see the text in Section~\ref{dendro.sec} for details). Each group is enclosed by its convex hull to distinguish itself from others. The red annotations in the top left-hand panel label the associations catalogued by \citet{Lucke1970}. The (0, 0) position corresponds to R.A.~(J2000)~=~05$^{\rm h}$37$^{\rm m}$40$^{\rm s}$, Dec.~(J2000)~=~$-$69$^{\rm o}$22$\arcmin$18$\arcsec$. The southwest blank corner corresponds to the region affected by worse signal-to-noise ratios associated with detector 16.}
\label{groups.fig}
\end{figure*}

The groups found at the first four levels with log($\Sigma_{\rm cr}\cdot$pc$^2$)~=~$-$1.7, $-$1.5, $-$1.3, and $-$1.1 are shown in Fig.~\ref{groups.fig}. To help distinguish different groups, each one is enclosed with its convex hull, shown as thick-lined polygons in the figure; 39 groups are found on the lowest level with log($\Sigma_{\rm cr}\cdot$pc$^2$)~=~$-$1.7. Most prominent are the two large groups, labeled ``A" and ``B" in Fig.~\ref{groups.fig}a, residing in the central part of the tile. Group~A consists of the 30~Dor complex, LH~100-LH~99-LH~90, and the diffuse LH~96, LH~111 to its south and east, which have already been mentioned in Section~\ref{distribution.sec}. On the other hand, Group~B is composed of LH~101-LH~104 and LH~103 in the star-forming regions N158 and N160, respectively. 48 groups are found on the second level with log($\Sigma_{\rm cr}\cdot$pc$^2$)~=~$-$1.5, and the groups are generally smaller in size compared with those on the first level. Group~A now splits into two large groups, corresponding to LH~100-LH~99-LH~90 and LH~96, a smaller LH~111 in the east, and several even smaller ones among them. Group~B also splits, with LH~101, LH~104, LH~103, and many other groups gaining their independence. 45 and 40 groups are found on the third and fourth levels, respectively. Results on higher levels are not shown here, but similar processes are continuously repeated, i.e., larger groups vanish and smaller groups emerge.

\begin{figure}
\center
\includegraphics[scale=0.75,angle=0]{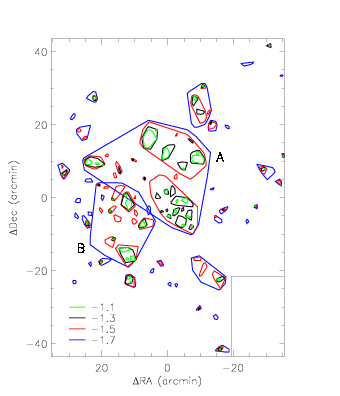}\\
\caption{Convex hulls enclosing the groups found with $N_{\rm min}^\star$~=~10 and critical surface densities of log($\Sigma_{\rm cr}\cdot$pc$^2$)~=~$-$1.7 (blue), $-$1.5 (red), $-$1.3 (black), and $-$1.1 (green). The (0, 0) position corresponds to R.A.~(J2000)~=~05$^{\rm h}$37$^{\rm m}$40$^{\rm s}$, Dec.~(J2000)~=~$-$69$^{\rm o}$22$\arcmin$18$\arcsec$. The southwest blank corner corresponds to the region affected by worse signal-to-noise ratios associated with detector 16.}
\label{convex.fig}
\end{figure}

\begin{figure*}
\center
\includegraphics[scale=0.75,angle=0]{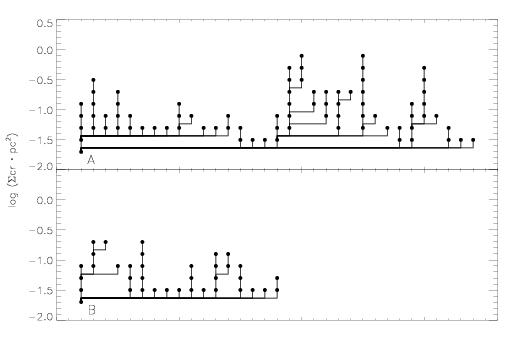}
\caption{Dendrograms of Group A (top) and Group B (bottom), illustrating the groups found at different levels and their connections. This figure only shows the results obtained with $N_{\rm min}^\star$~=~10 in the group-identifying algorithm (see the text in Section~\ref{dendro.sec} for details).}
\label{tree.fig}
\end{figure*}

To better display the relations between groups on different levels, we show all the convex hulls of groups found on the first four levels in Fig.~\ref{convex.fig}, and the dendrograms of Group A and Group B in Fig.~\ref{tree.fig}. One can, of course, construct a single dendrogram containing all groups found on all levels. Here we show only the dendrograms starting from Group A and Group B, which are its ``child dendrograms". Both Fig.~\ref{convex.fig} and Fig.~\ref{tree.fig} show that the stellar groups on lower levels very frequently split up into several smaller ones on higher levels. Thus, the stellar structures display a high degree of sub-grouping. This result agrees well with previous works, which also reveal hierarchically sub-grouped young stellar structures in individual star-forming regions \citep[e.g.][]{Schmeja2009, Kirk2011, Gouliermis2014} or over the whole galaxies \citep[e.g.][]{Gouliermis2010, Gouliermis2015, Gusev2014}.

\section{Group Sizes}
\label{size.sec}

\begin{figure}
\center
\includegraphics[scale=0.75,angle=0]{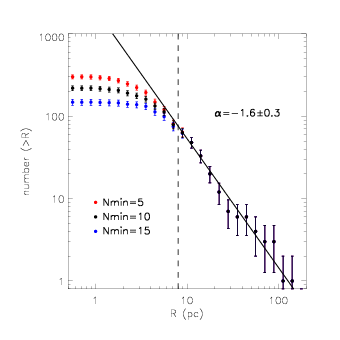}
\caption{Cumulative size distribution of stellar groups with Poissonian errorbars. The red, black and blue data points correspond to different values of $N_{\rm min}^\star$ used in the group-identifying algorithm.At large $R$, only the black points are seen because points of different $N_{\rm min}^\star$ overlap with each other.} The vertical dashed line shows the radius beyond which the stellar groups are complete. The solid line is the best-fitting power-law relation to the data points unaffected by incompleteness. Data points of different colors overlap at large sizes.
\label{size.fig}
\end{figure}

The stellar groups identified in Section~\ref{dendro.sec} have a wide range of sizes, from the largest Groups~A and B, which are more than 100~pc in extent, to the smallest ones on parsec scales. Fig.~\ref{size.fig} shows their cumulative size distribution, which includes stellar groups found on all levels of critical surface densities. The size of a group is estimated using
\begin{equation}
R = \dfrac{\sum_i |x_i - x_c| w_i}{\sum_i w_i}\ ,
\end{equation}
where $x_i$ is the position vector of the $i^{\rm th}$ group member, $w_i$ its weight assigned in the same way as described in Section~\ref{sample.sec} to account for the local photometric completeness, and $x_c$ the group center, which is calculated as
\begin{equation}
x_c = \dfrac{\sum_i x_i  w_i}{\sum_i w_i}\ .
\end{equation}
The cumulative size distribution is approximately a power law between $\sim$10~pc to $>$100~pc, with significant flattening at several parsecs.
As mentioned in Section~\ref{dendro.sec}, stellar groups are identified on levels of critical surface densities spaced by equal logarithmic steps of 0.2~dex; with smaller steps and thus more levels, more stellar groups can be included in obtaining the cumulative size distribution. We carried out a test by reducing this step to 0.1 and 0.05~dex and repeating the group identification process; the shapes of the resultant cumulative size distributions, and specifically the slopes of their power-law parts, are almost unchanged (not shown), suggesting that this does not affect the conclusion reached in this section. On the other hand, the flattening is due to incompleteness in our group-identifying algorithm. As mentioned in the previous section, smaller groups may be missed if they contain fewer UMS stars than the required minimum number of stars, $N_{\rm min}^\star$. Thus, we have calculated the cumulative size distribution multiple times for stellar groups identified with $N_{\rm min}^\star$~=~5, 10, and 15. As expected, fewer small-sized groups are identified with increasing values of $N_{\rm min}^\star$. Beyond $R$~=~8~pc, however, the cumulative size distribution remains almost unaltered. Thus we suggest that the stellar groups found by our group-identifying algorithm are complete above $R$~=~8~pc.

In the range not affected by incompleteness, the cumulative size distribution can be well fitted with a single power law with power-law index $\alpha$~=~$-$1.6~$\pm$~0.3 (fitting error). This indicates a scale-free behavior and agrees with the scenario of hierarchical star formation \citep{Elmegreen2000}. On the other hand, the cumulative size distribution for substructures inside a fractal follows
\begin{equation}
\label{size.eq}
N(>R) \propto R^{-D}\ ,
\end{equation}
where $N$ is the number of substructures with sizes larger than $R$, and $D$ is the fractal dimension \citep{Mandelbrot1983, Elmegreen1996}. Thus the UMS stars have a fractal dimension of $D_2$~=~1.6~$\pm$~0.3, by comparing the power-law cumulative group size distribution with Equation~\ref{size.eq}. Note $D_2$ is the fractal dimension of the UMS stars' $\textit{projections}$ on the two-dimensional plane perpendicular to the line of sight. It is not easy to recover the fractal dimension of the UMS stars' $\textit{volume}$ distributions, $D_3$, from $D_2$. \citet{Beech1992} suggested a relation $D_3$~=~$D_2$~+~1; however, this relation holds only if the perimeter-area dimension of a projected three-dimensional structure is the same as that of a slice \citep{Elmegreen2004}, and moreover, this relation is often shown to be unrealistic for star-forming regions \citep[e.g.,][]{Sanchez2005}. Based on simulations, \citet{Gouliermis2014} proposed a more general conversion between $D_3$ and $D_2$, which would suggest $D_3$~=~$D_2$~=~1.6 for this case; but this may still remain uncertain, since this real star-forming complex can be different in its three-dimensional structures from, and maybe also more complicated than, their simulated fractals.

The ISM also exhibits hierarchical structures of clumps and filaments on all scales \citep{Larson1981, Elmegreen2008}. These structures have power-law cumulative size distributions \citep[e.g.,][]{Elmegreen1996, RomanDuval2010}, very similar to the UMS stars as found here. Typical fractal dimensions reported are close to $D_3$~=~2.4; if $D_2$~=~$D_3 -1$~=~1.4 holds for the ISM, this value is consistent with that of the UMS stars, within the errors. This (at least qualitativaly) supports a scenario of hierarchical star formation, where the newly-born stars follow the the gas distribution \citep{Elmegreen2000}. For the origin of hierarchical structures in the ISM, the role of turbulence as the driving source is highlighted by recent studies \citep[e.g.,][]{Federrath2009}; as suggested by \citet{Elmegreen2000}, however, this is also possibly accompanied by agglomeration with fragmentation \citep{Carlberg1990, McLaughlin1996} as well as self-gravity \citep{deVega1996}.

Beyond the regime of binary and multiple systems on very small scales, the young stars in a few Galactic star-forming regions are hierarchically clustered with fractal dimensions $D_2$~=~1.4, as derived by \citet{Larson1995} for Taurus \citep[see also][]{Gomez1993}, or $D_2$~=~1.5, as estimated by \citet{Simon1997} for Taurus, Orion, and Ophiuchus. The projected fractal dimension we have derived agrees well with these studies. \citet{Gouliermis2014} reported bimodal stellar clustering for the SMC star-forming region, NGC~346; the auto-correlation function (ACF) of young stars is well described by a broken power law, with a shallower slope for scales smaller than $\sim$5.8~pc and a steeper one for scales above this value. The slope of the latter part corresponds a projected fractal dimension of $D_2$~=~1.4, also consistent with our result. They have shown that the ACF is best explained by two distinct stellar clustering components, with a centrally concentrated one dominant in the central part of the region, and an extended, hierarchical distribution across the observed field. In the case of 30~Dor-N158-N159-N160, however, while there may be centrally concentrated, non-hierarchical components on small scales (e.g., star clusters), their existence is not revealed by the cumulative size distribution shown here.

Significant for the intense star formation, the 30~Dor-N158-N159-N160 complex has been proposed to arise from galaxy interactions. \citet{deBoer1998} suggested that the star formation in this area may have been induced by the bow-shock formed as the LMC moves through the Milky Way's halo. Alternatively, \citet{Fujimoto1990} and \citet{Bekki2007} considered the interaction between the Magellanic Clouds. Despite these possible external influences, however, no significant difference in the value of $D_2$ is found compared with the star-forming regions mentioned above, given the measurement uncertainties.

On larger scales, young stars also exhibit galaxy-wide hierarchical distributions. Specifically, \citet{Elmegreen2001}, \citet{Elmegreen2006, Elmegreen2014} and \citet{Gouliermis2015} reported power-law cumulative size distributions of the young stellar groups in a sample of galaxies; in contrast, \citet{Bastian2007, Bastian2009} suggested that the young stellar groups in M33 and the LMC have log-normal size distributions. However, if we deem the flattening on small scales in their cumulative size distributions (their Figures~5~and~1, respectively) as caused by incompleteness, the large-scale parts can also be well described by single power laws. Note that \citet{Bastian2007, Bastian2009} required each of their identified stellar groups should contain at least 5 sources, which may miss small-sized groups; but they did not discuss the issue of incompleteness as has been done earlier in this section.

On the other hand, the reported projected fractal dimensions of the galaxy-wide hierarchies span a range from close to 1.0 to as large as 1.8 or more \citep{Elmegreen2001, Elmegreen2006, Elmegreen2014, Gouliermis2015}. $D_2$ derived in this work is close to those for NGC~628 \citep[$D_2$~=~1.5;][]{Elmegreen2006}, NGC~6503 \citep[$D_2$~=~1.7;][]{Gouliermis2015}, and part of the galaxy sample of \citet{Elmegreen2014}, but it can deviate more than the measurement uncertainties compared with other ones. \citet{Elmegreen2014} found that compared with large spiral galaxies or low surface brightness dwarfs, the starburst dwarfs or HII galaxies have larger projected fractal dimensions if they contain one or two dominant young stellar complexes; thus, the difference in $D_2$ may reflect different clustering under different conditions. Secondly, older stellar populations have a smaller degree of substructures due to dynamical evolutions \citep{Gieles2008, Bastian2009, Parker2014, Gouliermis2015}; this effect may also contributes to the difference in $D_2$, since these studies rely on different stellar samples which may be complicated mixtures of populations of various ages. Moreover, it is worth-mentioning that the derived fractal dimensions may also depend on the adopted method \citep{Federrath2009}, but in this work we do not attempt to explore this effect.

Using the same VMC tile as in this work, \citet{Romita2016} report the discovery of abundant embedded star clusters in molecular clouds. The size distribution of the embedded clusters spans the range of 0.25--2.25~pc with a peak at $\sim$1.1~pc, very different from ours. Through their visual inspections, they preferentially identified small and compact cluster candidates, which have high contrast of stellar surface density with respect to their surrounding areas. The group identification algorithm used in this work, however, does not introduce a preferential scale and reveals star groups over a continuous range of sizes. On the other hand, as mentioned above, the star groups identified with the relatively rare UMS stars are incomplete below $R$~=~8~pc. Thus, \citet{Romita2016} and this work focus on different subsections of the full hierarchy of young stellar structures. \citet{Bonatto2010} investigated the size distributions of star clusters and ``non-clusters" which are essentially nebular complexes and stellar associations. The former class has a steeper distribution, and the latter has a shallower distribution than ours. Their results are different from ours owing to two reasons. First, they have treated the two classes separately, while we do not try to distinguish star clusters, associations, or larger stellar structures. Second, they have used the catalog of extended objects of \citet{Bica2008}, which in turn was compiled from many previous works. Thus the effect of selection bias is hard to assess.

\section{Summary and Conclusions}
\label{summary.sec}

In this work we investigate the hierarchical stellar structures of the 30~Dor-N158-N159-N160 star-forming complex traced by young UMS stars observed with the VMC survey. We calculate the surface densities of the UMS stars, whose values cover two orders of magnitudes. We identify groups of UMS stars on different levels of critical surface densities. The larger-sized groups on lower levels often contain several smaller-sized ones on higher levels, and a high degree of sub-grouping is revealed by the dendrograms constructed to show the ``parent--child" relations between the groups on different levels. The stellar groups have sizes ranging continuously from several parsecs to more than 100~pc. Without characteristic sizes, the stellar groups have a power-law cumulative size distribution, with the power-law index indicating a projected fractal dimension $D_2$~=~1.6~$\pm$~0.3. We suggest that the cumulative size distribution of the UMS stars is related to their parental ISM in a scenario of hierarchical star formation, in which newly-born stars follow the gas distribution. The projected fractal dimension derived here is close to those reported for other star-forming regions as well as for several galaxies; the difference with respect to some other galaxies is also discussed.

\acknowledgements
We are grateful to the anonymous referee for the detailed and constructive comments on this paper. The analysis in this article is based on observations collected at the European Organisation for Astronomical Research in the Southern Hemisphere (ESO) under ESO program 179.B-2003. We thank the Cambridge Astronomical Survey Unit (CASU) and the Wide Field Astronomy Unit (WFAU) for providing calibrated data products under the support of the Science and Technology Facility Council (STFC) in the UK. We thank Jim~P.~Emerson for helpful suggestions. N.-C.S. and R.d.G. acknowledge funding support from the National Natural Science Foundation of China through grants 11373010, 11633005, and U1631102.


\begin{thebibliography}


\bibitem[Bastian et al.(2007)]{Bastian2007} Bastian, N., Ercolano, B., Gieles, M., et al.\ 2007, \mnras, 379, 1302
\bibitem[Bastian et al.(2009)]{Bastian2009} Bastian, N., Gieles, M., Ercolano, B., \& Gutermuth, R.\ 2009, \mnras, 392, 868 
\bibitem[Beech(1992)]{Beech1992} Beech, M.\ 1992, \apss, 192, 103 
\bibitem[Bekki \& Chiba(2007)]{Bekki2007} Bekki, K., \& Chiba, M.\ 2007, PASA, 24, 21
\bibitem[Bica et al.(2008)]{Bica2008} Bica, E., Bonatto, C., Dutra, C.~M., \& Santos, J.~F.~C.\ 2008, \mnras, 389, 678
\bibitem[Bolatto et al.(2000)]{Bolatto2000} Bolatto, A.~D., Jackson, J.~M., Israel, F.~P., Zhang, X., \& Kim, S.\ 2000, \apj, 545, 234
\bibitem[Bonatto \& Bica(2010)]{Bonatto2010} Bonatto, C., \& Bica, E.\ 2010, \mnras, 403, 996
\bibitem[Bressan et al.(2012)]{Bressan2012} Bressan, A., Marigo, P., Girardi, L., et al.\ 2012, \mnras, 427, 127 
\bibitem[Bressert et al.(2010)]{Bressert2010} Bressert, E., Bastian, N., Gutermuth, R., et al.\ 2010, \mnras, 409, L54 

\bibitem[Cambr{\'e}sy et al.(2013)]{Cambresy2013} Cambr{\'e}sy, L., Marton, G., Feher, O., T{\'o}th, L.~V., \& Schneider, N.\ 2013, \aap, 557, A29
\bibitem[Carlberg \& Pudritz(1990)]{Carlberg1990} Carlberg, R.~G., \& Pudritz, R.~E.\ 1990, \mnras, 247, 353
\bibitem[Cardelli et al.(1989)]{Cardelli1989} Cardelli, J.~A., Clayton, G.~C., \& Mathis, J.~S.\ 1989, \apj, 345, 245
\bibitem[Casertano \& Hut(1985)]{Casertano1985} Casertano, S., \& Hut, P.\ 1985, \apj, 298, 80 
\bibitem[Chabrier(2001)]{Chabrier2001} Chabrier, G.\ 2001, \apj, 554, 1274 
\bibitem[Cignoni et al.(2015)]{Cignoni2015} Cignoni, M., Sabbi, E., van der Marel, R.~P., et al.\ 2015, \apj, 811, 76
\bibitem[Cioni et al.(2011)]{Cioni2011} Cioni, M.-R.~L., Clementini, G., Girardi, L., et al.\ 2011, \aap, 527, A116 
\bibitem[Cohen et al.(1988)]{Cohen1988} Cohen, R.~S., Dame, T.~M., Garay, G., et al.\ 1988, \apjl, 331, L95 
\bibitem[Cross et al.(2012)]{Cross2012} Cross, N.~J.~G., Collins, R.~S., Mann, R.~G., et al.\ 2012, \aap, 548, A119

\bibitem[Dalton et al.(2006)]{Dalton2006} Dalton, G.~B., Caldwell, M., Ward, A.~K., et al.\ 2006, \procspie, 6269, 62690X
\bibitem[de Boer et al.(1998)]{deBoer1998} de Boer, K.~S., Braun, J.~M., Vallenari, A., \& Mebold, U.\ 1998, \aap, 329, L49
\bibitem[de Grijs et al.(2014)]{deGrijs2014} de Grijs, R., Wicker, J.~E., \& Bono, G.\ 2014, \aj, 147, 122
\bibitem[De Marchi et al.(2016)]{DeMarchi2016} De Marchi, G., Panagia, N., Sabbi, E., et al.\ 2016, \mnras, 455, 4373
\bibitem[de Vega et al.(1996)]{deVega1996} de Vega, H.~J., S{\'a}nchez, N., \& Combes, F.\ 1996, \nat, 383, 56

\bibitem[Efremov(1995)]{Efremov1995} Efremov, Y.~N.\ 1995, \aj, 110, 2757 
\bibitem[Elmegreen \& Falgarone(1996)]{Elmegreen1996} Elmegreen, B.~G., \& Falgarone, E.\ 1996, \apj, 471, 816
\bibitem[Elmegreen et al.(2000)]{Elmegreen2000} Elmegreen, B.~G., Efremov, Y., Pudritz, R.~E., \& Zinnecker, H.\ 2000, Protostars and Planets IV, 179 
\bibitem[Elmegreen \& Elmegreen(2001)]{Elmegreen2001} Elmegreen, B.~G., \& Elmegreen, D.~M.\ 2001, \aj, 121, 1507 
\bibitem[Elmegreen \& Scalo(2004)]{Elmegreen2004} Elmegreen, B.~G., \& Scalo, J.\ 2004, \araa, 42, 211
\bibitem[Elmegreen et al.(2006)]{Elmegreen2006} Elmegreen, B.~G., Elmegreen, D.~M., Chandar, R., Whitmore, B., \& Regan, M.\ 2006, \apj, 644, 879
\bibitem[Elmegreen(2008)]{Elmegreen2008} Elmegreen, B.~G.\ 2008, Mass Loss from Stars and the Evolution of Stellar Clusters, 388, 249
\bibitem[Evans et al.(2011)]{Evans2011} Evans, C.~J., Taylor, W.~D., H{\'e}nault-Brunet, V., et al.\ 2011, \aap, 530, A108
\bibitem[Elmegreen et al.(2014)]{Elmegreen2014} Elmegreen, D.~M., Elmegreen, B.~G., Adamo, A., et al.\ 2014, \apjl, 787, L15 
\bibitem[Evans et al.(2015)]{Evans2015} Evans, C.~J., Kennedy, M.~B., Dufton, P.~L., et al.\ 2015, \aap, 574, A13

\bibitem[Federrath et al.(2009)]{Federrath2009} Federrath, C., Klessen, R.~S., \& Schmidt, W.\ 2009, \apj, 692, 364
\bibitem[Fukui et al.(2008)]{Fukui2008} Fukui, Y., Kawamura, A., Minamidani, T., et al.\ 2008, \apjs, 178, 56-70
\bibitem[Fujimoto \& Noguchi(1990)]{Fujimoto1990} Fujimoto, M., \& Noguchi, M.\ 1990, \pasj, 42, 505

\bibitem[Galametz et al.(2013)]{Galametz2013} Galametz, M., Hony, S., Galliano, F., et al.\ 2013, \mnras, 431, 1596 
\bibitem[Gieles et al.(2008)]{Gieles2008} Gieles, M., Bastian, N., \& Ercolano, B.\ 2008, \mnras, 391, L93
\bibitem[Gieles \& Portegies Zwart(2011)]{Gieles2011} Gieles, M., \& Portegies Zwart, S.~F.\ 2011, \mnras, 410, L6 
\bibitem[Girardi et al.(2008)]{Girardi2008} Girardi, L., Dalcanton, J., Williams, B., et al.\ 2008, \pasp, 120, 583
\bibitem[Girichidis et al.(2012)]{Girichidis2012} Girichidis, P., Federrath, C., Allison, R., Banerjee, R., \& Klessen, R.~S.\ 2012, \mnras, 420, 3264
\bibitem[Gomez et al.(1993)]{Gomez1993} Gomez, M., Hartmann, L., Kenyon, S.~J., \& Hewett, R.\ 1993, \aj, 105, 1927
\bibitem[Gordon et al.(2003)]{Gordon2003} Gordon, K.~D., Clayton, G.~C., Misselt, K.~A., Landolt, A.~U., \& Wolff, M.~J.\ 2003, \apj, 594, 279
\bibitem[Gouliermis et al.(2010)]{Gouliermis2010} Gouliermis, D.~A., Schmeja, S., Klessen, R.~S., de Blok, W.~J.~G., \& Walter, F.\ 2010, \apj, 725, 1717
\bibitem[Gouliermis et al.(2014)]{Gouliermis2014} Gouliermis, D.~A., Hony, S., \& Klessen, R.~S.\ 2014, \mnras, 439, 3775
\bibitem[Gouliermis et al.(2015)]{Gouliermis2015} Gouliermis, D.~A., Thilker, D., Elmegreen, B.~G., et al.\ 2015, \mnras, 452, 3508
\bibitem[Gregorio-Hetem et al.(2015)]{GregorioHetem2015} Gregorio-Hetem, J., Hetem, A., Santos-Silva, T., \& Fernandes, B.\ 2015, \mnras, 448, 2504 
\bibitem[Gusev(2014)]{Gusev2014} Gusev, A.~S.\ 2014, \mnras, 442, 3711

\bibitem[Hony et al.(2015)]{Hony2015} Hony, S., Gouliermis, D.~A., Galliano, F., et al.\ 2015, \mnras, 448, 1847
\bibitem[Houlahan \& Scalo(1992)]{Houlahan1992} Houlahan, P., \& Scalo, J.\ 1992, \apj, 393, 172 

\bibitem[Irwin et al.(2004)]{Irwin2004} Irwin, M.~J., Lewis, J., Hodgkin, S., et al.\ 2004, \procspie, 5493, 411 

\bibitem[Johansson et al.(1998)]{Johansson1998} Johansson, L.~E.~B., Greve, A., Booth, R.~S., et al.\ 1998, \aap, 331, 857

\bibitem[Kennicutt \& Hodge(1986)]{Kennicutt1986} Kennicutt, R.~C., Jr., \& Hodge, P.~W.\ 1986, \apj, 306, 130 
\bibitem[Kerber et al.(2009)]{Kerber2009} Kerber, L.~O., Girardi, L., Rubele, S., \& Cioni, M.-R.\ 2009, \aap, 499, 697 
\bibitem[Kirk \& Myers(2011)]{Kirk2011} Kirk, H., \& Myers, P.~C.\ 2011, \apj, 727, 64
\bibitem[Krumholz(2014)]{Krumholz2014} Krumholz, M.~R.\ 2014, \physrep, 539, 49
\bibitem[Kutner et al.(1997)]{Kutner1997} Kutner, M.~L., Rubio, M., Booth, R.~S., et al.\ 1997, \aaps, 122, 255

\bibitem[Larson(1981)]{Larson1981} Larson, R.~B.\ 1981, \mnras, 194, 809 
\bibitem[Larson(1995)]{Larson1995} Larson, R.~B.\ 1995, \mnras, 272, 213
\bibitem[Lombardi et al.(2010)]{Lombardi2010} Lombardi, M., Lada, C.~J., \& Alves, J.\ 2010, \aap, 512, A67 
\bibitem[Lombardi et al.(2011)]{Lombardi2011} Lombardi, M., Alves, J., \& Lada, C.~J.\ 2011, \aap, 535, A16
\bibitem[Lucke \& Hodge(1970)]{Lucke1970} Lucke, P.~B., \& Hodge, P.~W.\ 1970, \aj, 75, 171

\bibitem[Mac Low \& Klessen(2004)]{MacLow2004} Mac Low, M.-M., \& Klessen, R.~S.\ 2004, Rev. Mod. Phys., 76, 125
\bibitem[McKee \& Ostriker(2007)]{McKee2007} McKee, C.~F., \& Ostriker, E.~C.\ 2007, \araa, 45, 565
\bibitem[McLaughlin \& Pudritz(1996)]{McLaughlin1996} McLaughlin, D.~E., \& Pudritz, R.~E.\ 1996, \apj, 457, 578
\bibitem[Mandelbrot(1983)]{Mandelbrot1983} Mandelbrot, B.~B.,\ 1983, The Fractal Geometry of Nature (San Francisco: Freeman)
\bibitem[Megeath et al.(2016)]{Megeath2016} Megeath, S.~T., Gutermuth, R., Muzerolle, J., et al.\ 2016, \aj, 151, 5

\bibitem[Nakajima et al.(2005)]{Nakajima2005} Nakajima, Y., Kato, D., Nagata, T., et al.\ 2005, \aj, 129, 776

\bibitem[Ott et al.(2008)]{Ott2008} Ott, J., Wong, T., Pineda, J.~L., et al.\ 2008, PASA, 25, 129 

\bibitem[Parker et al.(2014)]{Parker2014} Parker, R.~J., Wright, N.~J., Goodwin, S.~P., \& Meyer, M.~R.\ 2014, \mnras, 438, 620
\bibitem[Pietrzy{\'n}ski et al.(2013)]{Pietrzynski2013} Pietrzy{\'n}ski, G., Graczyk, D., Gieren, W., et al.\ 2013, \nat, 495, 76

\bibitem[Roman-Duval et al.(2010)]{RomanDuval2010} Roman-Duval, J., Jackson, J.~M., Heyer, M., Rathborne, J., \& Simon, R.\ 2010, \apj, 723, 492
\bibitem[Romita et al.(2016)]{Romita2016} Romita, K., Lada, E., \& Cioni, M.-R.\ 2016, \apj, 821, 51
\bibitem[Rosolowsky et al.(2008)]{Rosolowsky2008} Rosolowsky, E.~W., Pineda, J.~E., Kauffmann, J., \& Goodman, A.~A.\ 2008, \apj, 679, 1338-1351 
\bibitem[Rubele et al.(2012)]{Rubele2012} Rubele, S., Kerber, L., Girardi, L., et al.\ 2012, \aap, 537, A106 
\bibitem[Rubele et al.(2015)]{Rubele2015} Rubele, S., Girardi, L., Kerber, L., et al.\ 2015, \mnras, 449, 639 

\bibitem[S{\'a}nchez et al.(2005)]{Sanchez2005} S{\'a}nchez, N., Alfaro, E.~J., \& P{\'e}rez, E.\ 2005, \apj, 625, 849
\bibitem[Sabbi et al.(2012)]{Sabbi2012} Sabbi, E., Lennon, D.~J., Gieles, M., et al.\ 2012, \apjl, 754, L37
\bibitem[Schmeja et al.(2009)]{Schmeja2009} Schmeja, S., Gouliermis, D.~A., \& Klessen, R.~S.\ 2009, \apj, 694, 367
\bibitem[Schmeja(2011)]{Schmeja2011} Schmeja, S.\ 2011, Astronomische Nachrichten, 332, 172
\bibitem[Simon(1997)]{Simon1997} Simon, M.\ 1997, \apjl, 482, L81 
\bibitem[Sutherland et al.(2015)]{Sutherland2015} Sutherland, W., Emerson, J., Dalton, G., et al.\ 2015, \aap, 575, A25 

\bibitem[Tatton et al.(2013)]{Tatton2013} Tatton, B.~L., van Loon, J.~T., Cioni, M.-R., et al.\ 2013, \aap, 554, A33 

\bibitem[Walborn \& Blades(1997)]{Walborn1997} Walborn, N.~R., \& Blades, J.~C.\ 1997, \apjs, 112, 457 
\bibitem[Walborn et al.(2014)]{Walborn2014} Walborn, N.~R., Sana, H., Sim{\'o}n-D{\'{\i}}az, S., et al.\ 2014, \aap, 564, A40

\bibitem[Zaritsky et al.(2004)]{Zaritsky2004} Zaritsky, D., Harris, J., Thompson, I.~B., \& Grebel, E.~K.\ 2004, \aj, 128, 16061

\end{thebibliography}
\end{document}